\documentclass{epl}
\usepackage{amsmath}
\usepackage{graphicx}
\usepackage{amsfonts}
\usepackage{amssymb}

\institute{
  \inst{} Dept. of Material and Interfaces, Weizmann Institute of Science, 76100 Rehovot\\
}  {Phase
transitions: general studies }  {Biomolecules:
structure and physical properties }

\begin{document}

\title{Role of cross-links in bundle formation, phase separation and gelation of long filaments}
\author{A. G. Zilman\inst{}
\and S. A. Safran \inst{}}
\maketitle
\begin{abstract}
We predict the thermodynamic and structural behavior of solutions
of long cross-linked filaments. We
find that at the mean field level, the entropy of self-assembled
junctions induces an effective attraction between the filaments 
that can result in a phase separation into a connected network, in equilibrium with a dilute phase.
A connected network can also be formed in a non-thermodynamic
transition upon increase of the chain, or cross link density, or
with decreasing temperature. For rigid rods, at low temperatures,
we predict a transition from an isotropic network, to anisotropic
bundles of rods tightly bound by cross links, that is triggered by the
interplay between the configurational entropy of the cross-link
distribution among the rods, and the rotational and translational entropy of the
rods. 
\end{abstract}

\textit{Introduction} -- Branched structures and networks are
found in many physical, chemical and biological systems. Physical
examples include gels, microemulsions/wormlike micelles,
polymer-colloid dispersions and dipolar fluids
\cite{worm,porte,dip,flory,strey,talmon,mol,sack,water}. Gelation
and network formation have been studied in a number of
experimental \cite{talmon,strey,tanaka,porte,sack,worm} \ and
theoretical
\cite{drye,tanakat,rub,semenov,isaac,perc,elleuch,panizza} works
and simulations \cite{kumar,kindt}. However, the connection
between gelation and a first order phase separation that commonly
accompanies the network formation, remain elusive
\cite{rub,kumar,syneresis}. Gels of rigid filaments such as 
actin cross-lnked with $\alpha $- actinin \cite{sack}, or charged polyelectrolytes such as DNA, 
cross-linked with multivalent ions,
\cite{wong,bloomfield} show a qualitatively similar
thermodynamic behavior, and also have been shown to form compressed bundles of filaments, 
tightly bound by the cross-links. Their origin has been subject of recent
theoretical activity \cite{itamar0,itamar}.

\begin{figure}[ptb]
\onefigure [width=13 cm]{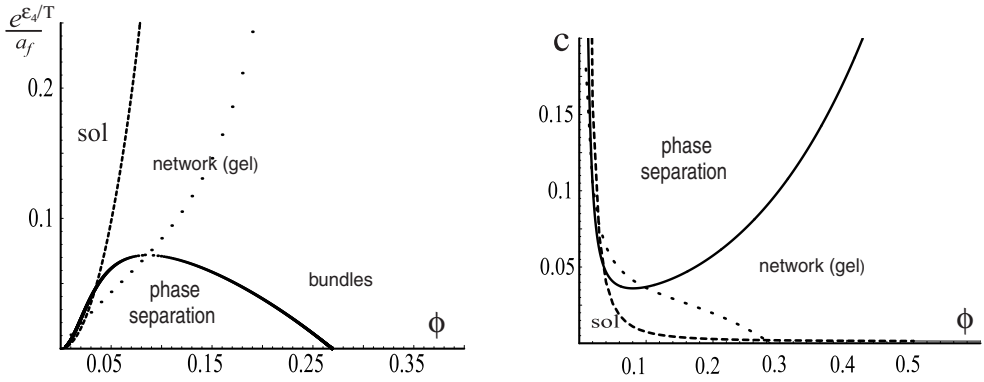} \caption{(a) The monomer
concentration-temperature, $\left(
\phi,e^{\epsilon_{4}/T}/a_{f}\right)  ,$ plane of the phase
diagram for strong, four-fold junctions for $\epsilon
_{j},\epsilon_{e}<0$ and $\epsilon_{4}/\epsilon_{e}=1$;
$c=0.05,\rho=0.005.$ The thick line shows the spinodal of the
junctions-ends transition. The dashed line is the percolation line
to the right of which a connected network is formed. The bundles
appear to the right of the dotted line; $a_{f}/q=1$. (b)
Monomer/crosslink concentration, $\left(  \phi,c\right)  $,
section of the phase diagram for $a_{f}e^{\epsilon_{4}/T}=0.015$,
$\left|  \epsilon _{e}/\epsilon_{4}\right|  =10/3,\rho=0.005.$ The
upper line delineates the region of the phase separation. The
lower, dashed, curve is the percolation line, above which a
connected network is formed. Bundles are predicted to
appear to the right of the dotted line.}%
\label{fig1}%
\end{figure}

In this paper, we predict, at the mean field level, the phase
diagram of a generic system of cross-linked equilibrium chains in
terms of two independent transitions in the system. The first is a
\emph{continuous}, non thermodynamic, topological transition,
where an infinite network spanning the entire volume of the system
is formed. A second, entropically driven, first-order
thermodynamic phase transition is predicted to occur even in the
absence of any specific interactions. For most values of the
control parameters, this latter transition is a sharp jump from a
dilute solution of weakly branched chains to a connected dense
network. \ The results, shown in Fig.\ref{fig1}, predict the
critical cross-link density and temperature for these
transitions.\ In addition, we find that the entropy of the cross
links can also drive a transition where the system forms
nematically aligned bundles of chains.\ \ The bundle transition,
as well as the phase separation and gelation lines predicted, are
consistent with the experiments on actin \cite{sack}.

We study a generic model system that consists of identical
monomers that self-assemble into chains, far above the
polymerization transition, where the chains just start to form
\cite{wheeler}. The chains, in turn, can branch in the presence of
cross links. In order to elucidate the universal nature of our
predictions, we assume \textit{no specific interactions} between
the monomers, except excluded volume, which means that the chains
are self-avoiding and mutually avoiding. The system is
characterized by the following parameters: the monomer density
$\phi$, the energies of the ends and junctions (branching points)
$\epsilon_{e}$ and $\epsilon_{j},$ measured relative to the energy
of a bond between two monomers within the chain, the temperature
$T$, the density of cross-links $c$, and the density of the
capping molecules responsible for the formation of the ends,
$\rho$. The numbers (per unit volume) of branching points,
$\phi_{j}$, and free ends, $\phi_{e},$ are not fixed but are
predicted as functions of $\phi,T$, $c$ and $\rho$. Here, we focus
on the case of sparse junctions and ends,
$\phi_{j},\phi_{e}\ll\phi$. The complementary case of very dense
junctions is better addressed using a different framework and is
studied elsewhere \cite{we}.

This formulation allows us to pinpoint the difference between weak and strong
cross-linking (i.e., physical gels vs. chemical gels) that correspond to
positive and negative junction energies $\epsilon_{j}$, respectively. In the
case of strong cross links, the density of the cross links must be
sufficiently low ($c\ll\phi)$ so that the total number of junctions formed is
small ($\phi_{j}<c\ll\phi)$. Of course, in the case of strong cross-linking
($\epsilon_{j}<0$), the experimental timescale at which true, equilibrium
thermodynamic behavior is observed, strongly increases as $\epsilon_{j}$ decreases.

\textit{Theoretical Model:\ }We begin by considering the system in
the grand-canonical ensemble where the network is in equilibrium
with a reservoir of \emph{ends }and \emph{junctions }of chemical
potential $\mu_{e}$ and $\mu_{j}$ respectively. In this
formulation the numbers of ends and junctions are not conserved
but determined by the values of $\mu_{e}$ and $\mu_{j}$,
respectively; these are calculated below. Both junctions and ends
can be viewed as ``thermal defects'' of the system whose ``ground
state'' we consider to be an assembly of infinite linear chains.
For flexible chains, in the random mixing approximation
\cite{flory}, the excluded volume part of the free energy is
$\frac{1}{2}\phi^{2}$. For rigid rods, the angular average of the
excluded volume interaction between the rods is
$\frac{1}{2}vn^{2}$, with $v\simeq dL^{2},$ $n=\phi/L$ , where $d$
is the rod diameter (taken as unity in our lattice model) and $L$
is the average length of the chains \cite{onsager,prost}. The
excluded volume interaction thus scales as $\frac {1}{2}\phi^{2}$
for both flexible chains and rigid rods.

General thermodynamic considerations (as well as rigorous treatments
\cite{big}) show that in the limit of a low density of non-interacting ends
and junctions, every non-conserved, thermally generated junction or end lowers
the free energy by $k_{\text{B}}T$. At the mean field level of analysis, the
grand-canonical potential of the system (per unit volume) is given by
$\Omega(\phi,\mu_{j},\mu_{e},\epsilon_{j},\epsilon_{e})=\frac{1}{2}\phi
^{2}-\phi_{e}-\phi_{j}$ where $\phi_{e}$ and $\phi_{j}$ are the densities of
ends and junctions. The physics of the problem lies in the dependence of
$\phi_{e}$ and $\phi_{j}$ on the parameters of the system, such as the monomer
density $\phi$, temperature $T$ and the chemical potentials $\mu_{j}$ and
$\mu_{e}$. It can be understood via a simple probabilistic argument: two ends
are formed whenever a bond is broken by thermal fluctuations in the presence
of two available end cap molecules. To the first order in the number of ends
and junctions, the number of bonds is proportional to the number of the
monomers, $\phi$. Taking into account the free energy of formation of two
ends, $2(\epsilon_{e}-\mu_{e})$, the probability of bond breaking is $\phi
e^{2(\mu_{e}-\epsilon_{e})/T}$. In equilibrium, this must be equal to the
probability of bond formation from the coalescence of two ends. \ In the
random mixing approximation, the probability of a collision of two ends is
$\phi_{e}^{2}a_{1}^{-2}$ \ where the numerical prefactor $a_{1}$ reflects the
microscopic properties of the ends, such as flexibility of the chains and
effective collision cross-section.\ We thus find $\phi_{e}=a_{1}e^{(\mu
_{e}-\epsilon_{e})/T}\phi^{1/2}$. Similarly, an $f$-fold junction can form
through a collision of $f-2$ ends and an internal monomer (or a collision of
$f$ ends). The creation of $f-2$ ends at the expense of a junction costs an
energy $(f-2)(\epsilon_{e}-\mu_{e})-\epsilon_{j}+\mu_{j}+T\ln a_{f}$, where
the coefficient $a_{f}$ reflects the microscopic degrees of freedom of the
junction, and includes the entropy of configurations of the bonds and monomers
in the junction. The probability of junction break-up is thus equal to
$\phi_{j}a_{f}{-1}e^{\left(-\mu_{j}+\epsilon_{j}\right)/T}e^{(f-2)(-\epsilon
_{e}+\mu_{e})/T}$. Equating this with the probability of collision of $f-2$
ends with an internal monomer, $\phi_{e}^{f-2}\phi a_{1}^{2-f}$, we find for
the junction and end densities
\begin{equation}
\phi_{j}=a_{f}e^{\left(  \mu_{j}-\epsilon_{j}\right)  /T}\phi^{f/2}\qquad
\phi_{e}=a_{1}e^{\left(  \mu_{e}-\epsilon_{e}\right)  /T}\phi^{1/2}\nonumber
\end{equation}
Therefore, in the random mixing approximation, the grand canonical potential
of the system for the case of sparse junctions and ends ($\phi_{j},\phi_{e}%
\ll\phi$) is :
\begin{equation}
\Omega(\phi,\mu_{j},\mu_{e})/T=\frac{1}{2}\phi
^{2}-a_{1}e^{(\mu_{e}-\epsilon_{e})/T}\phi^{1/2}-a_{f}e^{(\mu_{j}-\epsilon
_{j})/T}\phi^{f/2}\label{fmu}%
\end{equation}
Subsequently, we put $a_{1}=1$ for convenience. Equation
(\ref{fmu}) can also be derived rigorously
\cite{panizza,isaac,big} as described in \cite{big}, or, in some
particular cases, by heuristic methods \cite{act,drye,kindt}.

It is important to note that at this mean field level it is of no
consequence whether in the absence of any junctions or ends (zero
temperature) the ``ground state'' consists of infinite chains or
closed rings. For strong cross-linking, the junctions persist at
low temperatures and the number of closed linear rings is
exponentially small compared with number of branched aggregates.
Even for weak junctions, the number of closed rings relative to
the number of chains is smaller by a factor $L^{-3/2}$ (where $L$
is the chain length).\ Their influence is of importance only at
extremely low temperatures and densities,
Ref.\cite{semenov,wheeler}. In addition, the role of intra-cluster
loops \cite{kindt} is outside the scope of our model, because the
information about the long range correlations along a chain is
lost in the mean field approximation.

We now proceed to transform the grand-canonical potential
$\Omega(\phi,\mu _{j})$ of Eq.($\ref{fmu}$) and express the free
energy as a function of the physical cross link and end cap
densities, $c$ and $\rho,$ respectively, instead of the chemical
potentials $\mu_{j}$ and $\mu_{e}$. The free energy of a system of
branched chains in terms of the junction and ends densities
$\phi_{j}$ and $\phi_{e},$ is given by the Legendre transform of
the free energy Eq. $\left(  \ref{fmu}\right)  $
$F(\phi,\phi_{j},\phi_e)=\Omega(\phi,\mu
_{j},\mu_e)+\mu_{j}\phi_{j}+\mu_{e}\phi_{e}$:
\begin{equation}
F(\phi,\phi_{j},\phi_{e})/T=\frac{1}{2}\phi^{2}+\phi_{j}(\ln\phi_{j}-1)+\phi_{e}\left(
\ln\phi_{e}-1\right)  +\phi_{j}(\epsilon_{j}/T-\ln
a_{f})+\phi_{e}\epsilon
_{e}/T-\frac{f}{2}\phi_{j}\ln\phi-\frac{1}{2}\phi_{e}\ln\phi\label{fj}%
\end{equation}
The first term is the excluded volume of the chains, the next four
terms represent the free energy of an ``ideal gas'' of junctions
and ends, and the last two terms are the increase in the free
energy of this ``ideal gas'' due to the fact that the junctions
and the ends are constrained to a network. Each junction confines
$f/2$ monomers to the same point in space which increases the free
energy by $\frac{f}{2}\phi_{j}\mu_{\phi}$ where $\mu_{\phi}\simeq
T\ln\phi$ is the chemical potential of a ``gas'' of monomers.
Similarly, last term reflects the reduction in the entropy of a
''gas'' of ends due to the fact that two ends are constrained to
lie on the same chain. Because each junction requires one
cross-link molecule and each end uses up one end cap molecule, the
$total$ free energy of the system is $F^{\text{tot}}(\phi
,\phi_{j},c)=F(\phi,\phi_{j},\phi_e)+F_{c}(c-\phi_{j})+F_{e}\left(
\rho-\phi _{e}\right)  $, where $F_{c}(c-\phi_{j})$ and
$F_{e}\left(  \rho-\phi _{e}\right)  $\ \ are the free energies of
the unbound cross-links and ends, respectively. We take these both
to be ideal solutions: $F_{e}$($\psi )=F_{c}\left(  \psi\right)
=T\psi(\ln\psi-1)$. The total free energy
$F^{\text{tot}}(\phi,\phi_{j},\phi_{e},c,\rho)$ is minimized with
respect to $\phi_{j}\,$\ and $\phi_{e}$ to find the equilibrium
end and junction density as a function of
$\phi,\epsilon_{j},\epsilon_{e},c,\rho$ and $T$. This predicts
that the junction and end densities vary with the monomer,
cross-link and end-cap densities as
\begin{equation}
\phi_{e}=\rho e^{-\epsilon_{e}/T}\phi^{1/2}/\left(  1+e^{-\epsilon_{e}/T}%
\phi^{1/2}\right)  \qquad\qquad\phi_{j}=ca_{f}e^{-\epsilon_{j}/T}\phi
^{f/2}/\left(  1+a_{f}e^{-\epsilon_{j}/T}\phi^{f/2}\right)  \label{fij}%
\end{equation}
It is important to note that the density of junctions $\phi_{j}$ behaves
rather differently for strong ($\epsilon_{j}<0$) and weak $\left(
\epsilon_{j}>0\right)  $ cross-linking. In the limit of very strong junctions
($\epsilon_{j}<0,|\frac{\epsilon_{j}}{T}|\gg1)$, the junction density
saturates to a value equal to the cross-link concentration, $c$.

This means, that in this limit $\left(  i\right)  $ almost all cross-link
molecules are found in the junctions $\left(  ii\right)  $ the junctions
persist at low temperatures. In the opposite limit of weak junctions,
$\phi_{j}\simeq ca_{f}e^{-\epsilon_{j}/T}\phi^{f/2}\ll c$ which means that
$\left(  i\right)  $ most of the cross-link molecules are found in solution
and not in junctions $\left(  ii\right)  $ at low temperatures, the number of
thermally generated junctions tends to zero. Also, it follows from
Eq.(\ref{fij}) that the condition of sparse junctions, used in the derivation
of Eq.(\ref{fmu}) is always satisfied as long as $c\ll\phi$.

Substituting the expression of Eq.(\ref{fij}) for $\phi_{j}$ back into the
free energy $F,$ gives:
\begin{equation}
F^{\text{tot}}(\phi,c)=\frac{1}{2}\phi^{2}+c\left(  \ln c-1\right)
+\rho\left(  \ln\rho-1\right)  -c\ln(1+a_{f}e^{-\epsilon_{j}/T}\phi
^{f/2})-\rho\ln\left(  1+e^{-\epsilon_{e}/T}\phi^{1/2}\right)  \label{ftot01}%
\end{equation}
The last two terms are the free energy reduction due to the presence of
thermally generated ends and junctions. Although both these terms represent
negative contributions to the free energy, the term associated with the ends
is thermodynamically equivalent to a \emph{repulsion} while the term
associated with the junctions induces an effective thermodynamic
\emph{attraction} between the monomers.\textit{\ }This can be seen by
considering the osmotic pressure, $\Pi=\rho\frac{\partial F}{\partial\rho
}+\phi\frac{\partial F}{\partial\phi}+c\frac{\partial F}{\partial c}-F$, to
which the junctions give a \textit{negative} contribution, $-\frac{f}{2}%
ca_{f}e^{-\epsilon_{j}/T}\phi^{f/2}/(1+a_{f}e^{-\epsilon_{j}/T}\phi^{f/2})$
while the ends give a \emph{positive }\ contribution $+\frac{1}{2}\rho
e^{-\epsilon_{e}/T}\phi^{1/2}/(1+e^{-\epsilon_{e}/T}\phi^{1/2})$.

\textit{Phase Separation and Gelation:\ }If the density of cross
link molecules is sufficiently high, or the temperature is
sufficiently low, the junction induced attraction can be strong
enough to drive a phase separation between a dense, junction rich,
network and a dilute phase consisting either of weakly branched
chains or a very sparse network.

For thermodynamic stability, the free energy $F^{\text{tot}}(\phi,c,\rho)$
must be convex with respect to all the thermodynamic variables $\phi$, $c$ and
$\rho$. Technically, this requires that the matrix of second derivatives of
$F^{\text{tot}}(\phi,c,\rho)$ is positive definite (i.e., has three positive
eigenvalues). This condition defines the spinodal surface of the phase
separation in the phase space of the system, and the critical density of
cross-links, $c_{\text{s}}(\phi,\rho,T),$ as a function of  the  end cap
concentration, the monomer concentration and the temperature, is given by:
\begin{equation}
c_{\text{s}}(\phi,\rho,T)=\frac{4}{f\left(  f-2\right)  a_{f}}\left(
\frac{\phi^{(4-f)/2}e^{\epsilon_{j}/T}}{1-\phi}+\frac{1}{4}\rho\frac
{\phi^{(1-f)/2}e^{(\epsilon_{j}-\epsilon_{e})/T}}{1+\phi^{1/2}e^{-\epsilon
_{e}/T}}\right)  \left(  1+a_{f}e^{-\epsilon_{j}/T}\phi^{f/2}\right)
\label{cspin}%
\end{equation}
\ In the $\left(  c,\phi\right)  $ plane, the system is stable for
cross-link densities $c$ $<c_{\text{s}}(\phi)$, while for
$c>c_{\text{s}}(\phi)$ the system separates into a dense network
with many junctions, in equilibrium with a dilute phase.
Similarly, in the $\left(  T,\phi\right)  $ plane, at a given
cross-link density $c$, the system is unstable for a certain range
of monomer densities $\phi $ as shown in Fig. 1. The dilute phase
can be either a very sparse network or consist of disjointed,
weakly branched clusters. The dense phase is usually a connected
network, as discussed below. The thermodynamic behavior is rather
different for weak and for strong cross-links. In the case of weak
cross-links ($\epsilon_{j}>0$, $\epsilon_{j}/T\gg1)$ the
junction-induced attraction is strong enough to drive the phase
separation only for three-fold junctions, $f=3$. \ In addition,
the phase transition in the case of weak junctions is
\emph{reentrant }because the number of thermally generated
junctions tends to zero as the temperature is lowered as follows
from Eq.$\left( \ref{fij}\right) $. In contrast, in the case of
strong junctions ($\epsilon_{j}<0$), the phase separation occurs
for junctions of any functionality as follows from
Eq.($\ref{cspin})$; the transition is not reentrant because the
number of junctions \emph{increases }with decreasing temperature,
as follows from Eq.$\left(  \ref{fij}\right)  .$ \  The predicted
phase transition is \textit{entropic in origin}. In the dense
phase, the entropy of the large number of the formed junctions
compensates for the loss of the translational entropy of the
monomers.

For sparse junctions, in the mean field approximation
\cite{flory,isaac,big}, a connected network spanning the entire
system is formed when $f\left( f-2\right)  \phi_{j}=\phi_{e}$ .
This gelation or percolation transition, is a continuous, purely
\emph{topological} transition that has no thermodynamic signature.
Thus, a connected network spanning the entire system is formed
when the cross link concentration exceeds the value
\[
c_{\text{perc}}= \rho /(f(f-2)a_{f}) \phi^{\left(  1-f\right)
/2}e^{(\epsilon_{j}-\epsilon_{e})/T}\left(  1+e^{-\epsilon_{j}/T}\phi
^{1/2}\right)  /\left(  1+a_{f}e^{-\epsilon_{e}/T}\phi^{f/2}\right)
\text{\ \ \ }%
\]
This allows one to predict the gelation line in the phase diagram
of the system, as shown in Fig. 1.

\textit{In vitro} experiments on solutions of rigid actin filaments,
cross-linked with $\alpha$-actinin and in presence of the capping protein
gelsolin, found that the system undergoes a continuous, non-thermodynamic
transition from an entangled network of branched clusters to a ``microgel''
\cite{sack} that results in a sharp increase in viscosity, and can be
interpreted in terms of percolation. At lower temperatures, it was found that
the system undergoes a first-order phase separation into a dilute solution of
short chains in equilibrium with a dense network that consists mostly of
bundles of actin filaments, tightly bound by cross-links. These observations
are consistent with the phase diagram of Fig. 1.

\textit{Bundles:\ }The formation of bundles can be understood in
terms of the entropy of rigid chains (rods) and cross links that
join two rods or four monomers, $f=4$. In the bundle phase, the
rotational and translational entropy of the rods is lower than in
the isotropic, network phase. However, the entropy of positioning
the junctions on the parallel ``tracks'' formed by the rods found
in bundles is higher than in the network, where junction positions
are confined to the intersections between the rods. It turns out
that this effect favors the formation of bundles at \ low
temperatures. As a first step, we estimate that the transition
from an isotropic network to bundles takes place when their free
energies become equal. Assuming that the energy of the ends is
large and negative, so that all the end cap molecules are bound to
chain ends, the free energy of the isotropic phase (neglecting a
term that depends on the number of end caps and is the same in
both phases) is given by Eq.\ref{ftot01}:
\[
F_{i}/T\simeq\frac{1}{2}\phi^{2}+c(\ln c-1)-c\ln\left(  1+a_{f}e^{-\epsilon
_{j}/T}\phi^{2}\right)  +\rho\left(  \ln\left(  \rho/\phi^{1/2}\right)
-1\right)
\]

We describe the bundle state as a collection of $N_{b}$ bundles of $m$ rods of
length $L$.\ \ The $N_{j}$ junctions formed by the cross links can be placed
between any two adjacent rods. If the number of nearest neighbor rods in a
bundle is $2q$, there are $N_{b}(qm-qm^{1/2}/2)$ available one-dimensional
``tracks'' for the junctions, because the rods at the perimeter of a bundle
have half as many neighbors as the interior rods; the number of\ perimeter
rods $\sim m^{1/2}$ for large bundles. The junction entropy is then $\sim
\frac{1}{N_{j}!}\left(  qN_{b}Lm\left[  1-1/\left(  2m^{1/2}\right)  \right]
\right)  ^{N_{j}}.$ Noting that $N_{b}Lm$ is the total number of monomers in
the system and recalling that the energy of a junction is $\epsilon_{j}$, we
find for the free energy of the bundle phase
\[
F_{b}/T\simeq\phi_{j}\left(  \ln\phi_{j}-1\right)  -\phi_{j}\ln\left(
q\phi\left(  1-1/\left(  2m^{1/2}\right)  \right)  \right)  +\phi_{j}%
\epsilon_{j}/T+\left(  c-\phi_{j}\right)  \left(  \ln\left(  c-\phi
_{j}\right)  -1\right)
\]
where the last term accounts for the translational entropy of the free
cross-links in the solution. Note that the junction-related part of the bundle
state is \emph{lower }than the corresponding part of the free energy of an
isotropic network, eq.(\ref{fj}), because the entropic cost of confining a
junction to lie on a track between two parallel rods is $-\ln\left(
q\phi\left(  1-1/\left(  2m^{1/2}\right)  \right)  \right)  $ and is lower
than the corresponding cost of confinement to an intersection\ of two rods in
a three dimensional network, $-2\ln\phi$. Minimizing the free energy as a
function of the junction density, $\phi_{j},$ gives for the junction-related
part of the free energy, $F_{b}/T\simeq c\left(  \ln c-1\right)  -c\ln\left(
1+e^{-\epsilon_{j}/T}q\phi\left(  1-1/\left(  2m^{1/2}\right)  \right)
\right)  $.

In addition to the configurational entropy of cross-link
distribution among the filaments, the free energy of the bundled
states includes the following contributions: $\left( i\right) $
the translational entropy of the bundles $\frac{\rho}{m}\left(
\ln\frac{\rho}{m\phi^{1/2}}-1\right)  $ and $\left(  ii\right)  $
the \emph{reduction }in the rotational entropy of the bundles
relative to the isotropic phase, $\rho\left(  1-\frac{1}{m}\right)
\ln4\pi$ \cite{prost,onsager,itamar0} and $\left(  iii\right)  $
the excluded volume repulsion between the bundles,
$\frac{1}{2}v\left( \phi/(mL)\right) ^{2}$ where $v\simeq
m^{1/2}L^{2}$ \cite{onsager}, because the diameter of the bundle
of $m$ rods is proportional to $m^{1/2}$. Without (i) and (ii),
that balance the increase of the configurational cross-link
entropy in the bundle phase, the bundles would be always preferred
thermodynamically. Although the free energy of the bundle state
can have a minimum at finite values of $m$, it turns out that this
region is small \cite{act}, and for a first estimate we assume
that the bundle state consists of one macroscopic bundle. In the
limit of $m\rightarrow\infty$, $F_{b}/T\simeq c\left(  \ln
c-1\right) -c\ln\left( 1+e^{-\epsilon_{j}/T}\phi\right)
+\rho\ln4\pi$. The transition to bundles, therefore, takes place
when $F_{b}<F_{i},$ that is
$e^{\epsilon_{j}/T}/a_{f}<\phi\frac{qe^{A}/a_{f}-\phi}{1-e^{A}}$
where $Ac=\frac{1}{2}\phi^{2} -\rho\ln4\pi$, as shown
in Fig.1. The location of the network to bundle transition line in
the phase diagram is sensitive to the value of the parameter
$q/a_{f}$, where $a_{f}$ reflects the number of internal
configurations of a cross-link in the isotropic network relative
to parallel bundles; increasing $q/a_{f}$ decreases the threshold
for bundle formation. The parameter $a_{f}$, as well as the
energies $\epsilon_{e}$ and $\epsilon_{j}$ can be estimated
experimentally from the number of bound cross-links, and the
length distribution of chain segments among them.\ However, when
comparing with the experimental results of \cite{sack}, one should
bear in mind, that for long chains, the formation of the bundles
might be kinetically inhibited and may not be experimentally
observable.

\acknowledgments We thank I. Borukhov, J. Isaacson, J. Kindt, T. Lubensky, E.
Sackmann, A. Semenov, R. Strey and P. Pincus for useful discussions. The
authors gratefully aknowledge support of the Donors of the Petroleum Research
Fund of the ACS, the German-Israel Foundation, and the Schmidt Minerva Center.


\begin{thebibliography}{99}
\bibitem{mol}M. Albers \emph{et al.}, \textit{Molecular Biology of the Cell},
Garland Publishing, New York, 1994

\bibitem{flory}P. J. Flory, Principles of Polymer Chemistry, Cornell
University Press, 1981.

\bibitem{water}D. Eizenberg, W. Kaufmann, The structure and Properties of
Water, Oxford University Press, 1969

\bibitem{sack}M. Tempel, G. Isenberg, E. Sackmann, Phys. Rev. E, 54, 1802 (1996).

\bibitem{dip}T. Tlusty, S.A. Safran, Science, 290, 1328 (2000)

\bibitem{wheeler}J.C. Wheeler, P. Pfeuty, Phys. Rev. A, 24, 1050 (1981)

\bibitem{elleuch}F. Lequeux, K. Elleuch, P. Pfeuty, J. Phys I, 5, 465 (1995)

\bibitem{rub}A.N. Semenov, M. Rubinstein, Macromolecules, 31, 1373 (1988)

\bibitem{kumar}S.K. Kumar, A. Z. Panagiotopoulos, Phys. Rev. Lett., 82, 5060
(1999) \textit{and references}

\textit{\ therein}

\bibitem{syneresis}H. Takeshita \emph{\ et.al.} Macromolecules, 34 , 7894
(2001) \emph{\ \ and referneces therein}

\bibitem{kindt}J. T. Kindt, J. Phys. Chem. B, 106, 8223 (2002)

\bibitem{isaac}T.C. Lubensky, J. Isaacson, Phys. Rev. Lett. 41, 829 (1978);
T.C. Lubensky, J. Isaacson, Phys. Rev. A 20, 2130-2146 (1979)

\bibitem{perc}A.B. Harris $et$ $al.,$ Phys. Rev. Lett., 35, 327 (1975); R.
Kikuchi, J. Chem. Phys, 53, 2713 (1970); A. Coniglio, Phys. Rev. B, 13, 2194
(1976); K.K. Murata, J. Phys. A, 12, 81 (1979)

\bibitem{tanakat}F. Tanaka, Physica A, 257, 245 (1998), F. Tanaka,
Macromolecules, 31, 384 (1998)

\bibitem{porte}M. Filali \emph{et. al.}, J. Phys. Chem. B. 105, 10528 (2001)

\bibitem{onsager}L. Onsager, Ann. N. Y. Acad. Sci., 51, 627 (1949)

\bibitem{prost}P.-G. DeGennes, J. Prost, The Physics of Liquid Crystals,
Oxford University Press, 1995

\bibitem{semenov}A. N. Semenov, I. A. Nyrkova, M. E. Cates, Macromolecules,
28, 7879 (1995)

\bibitem{drye}T. Drye, M.E. Cates, J. Chem. Phys. 96 (2), 1367 (1992); T.
Tlusty T, S.A. Safran, R. Strey, Phys. Rev. Lett., 84, 1244 (2000)

\bibitem{itamar0} I. Borukhov et. al. Phys. Rev. Lett., 87, 158101 (2001)

\bibitem{wong} J. C. Butler, T. Angelini, J. X. Tang, G. C. L. Wong,
Phys. Rev. Lett., 91, 028301 (2003) \textit{and references therein}

\bibitem{bloomfield}V. A. Bloomfield, Biopolymers, 44, 269 (1997)

\bibitem{shklovskii}T. T. Nguyen, I. Rouzina, B. I. Shklovskii,
 J Chem. Phys. 112, 2562 (2000)

\bibitem{talmon}A. Berheim-Groswasser $et$ $al$., Langmuir, 15: 5448 (1999);
A. Berheim-Groswasser $et$ $al$., Langmuir, 16, 4131 (2000)

\bibitem{strey}R. Strey $et$ $al.$, J. Chem. Phys, 105, 1175 (1996); F.
Lichterfeld $et$ $al$, J. Phys. Chem., 90, 5762 (1986)\textit{\ and references therein}

\bibitem{worm}A. Khatory $et$ $al.,$ Langmuir, 9, 933 (1993); E. Buhler $et $
$al.,$ J. Phys II, 5, 765 (1995)

\bibitem{tanaka}T. Tanaka $et$ $al.,$ Phys. Rev. Lett, 42, 1556 (1979); T.
Miura $et$ $al.,$ Phys. Rev. E, 54, 6596 (1996); A. Moussaid \emph{et. al.},
J. Phys II, 1, 637 (1991) \textit{and references therein}

\bibitem{panizza}G. Cristobal \emph{et al.}, Physica A, 268, 50 (1999)

\bibitem{big}A.G. Zilman and S.A. Safran, Phys. Rev. E. 66, 051107 (2002)

\bibitem{we}A. Zilman, J. Kieffer, F. Molino, G. Porte, S. Safran, Phys. Rev. Lett., 91, 015901,(2003)

\bibitem{itamar} I. Borukhov, K. C. Lee, R. F. Bruinsma, W. M. Gelbart \textit{et. al.},
J Chem. Phys., 117, 462 (2002)

\bibitem{act}A. Zilman, S. Safran, \emph{to be published}

\end{thebibliography}
\end{document}